\documentstyle[prl,aps,epsfig]{revtex}
\begin{document}
\twocolumn[\hsize\textwidth\columnwidth\hsize\csname@twocolumnfalse\endcsname

\title{Novel Josephson effects between multi-gap and single-gap superconductors}
\author{D.F. Agterberg$^{1,4}$, Eugene Demler$^{2}$, and B. Janko$^{3,4}$}
\address{$^1$ Department of Physics, University of Wisconsin-Milwaukee, Milwaukee, WI 53201}
\address{$^2$ Department of Physics, Harvard University, Cambridge, MA 01238}
\address{$^3$ Department of Physics, University of Notre Dame, Notre Dame, IN 46617}
\address{$^4$Materials Science Division, Argonne National Laboratory,
9700 South Cass Avenue, Argonne, IL 60439}
\maketitle
\begin{abstract}
Multi-gap superconductors can exhibit qualitatively new phenomena due to existence of multiple order parameters. Repulsive electronic interactions
may give rise to a phase difference of $\pi$ between the phases of the order parameters. Collective modes due to the oscillation of the relative
phases of these order parameters are also possible. Here we show that both these phenomena are observable in Josephson junctions between a
single-gap and a multi-gap superconductor. In particular, a non-monotonic temperature dependence of the Josephson current through the junction
reveals the existence of the $\pi$ phase differences in the multi-gap superconductor. This mechanism may be relevant for understanding several
experiments on the Josephson junctions with unconventional superconductors. We also discuss how the presence of the collective mode resonantly
enhances the DC Josephson current when the voltage across the junction matches the mode frequency. We suggest that our results may apply to
MgB$_2$, $2H$-NbSe$_2$, spin ladder and bilayer cuprates.
\end{abstract} \pacs{PACS numbers:74.50.+r,74.20.Mn}]

\narrowtext

Multi-band superconductors have been the subject of theoretical investigation since the original work of Suhl, Matthis, and Walker \cite{suh59}.
Experimentally, it has recently been shown that MgB$_2$ \cite{nag01,bou01,sza01,che01,tsu01,giu01} and $2H$-NbSe$_2$ \cite{yok01} belong to this
class.
In this article we examine some qualitatively new features associated
with superconductivity in the multi-band
materials. We will focus on examining Josephson junctions between a
multi-band superconductor and a single band superconductor. We show
that such a junction can reveal important information about the role
of electronic interactions in the pairing mechanism and further be
used to detect superconducting collective modes specific to multi-band
superconductors.

A phase difference of $\pi$ between the gap on different bands occurs when repulsive electronic interactions between the different bands play an
important role in creating the superconducting state \cite{kon63,dag96,nor99,ima01,yam01}. Such a mechanism has been argued to be relevant for the
spin-ladder cuprate superconductors \cite{sch01-4}, with the gap being of opposite sign on the bonding and antibonding bands of the two legs of the
ladder  \cite{dag96,nor99}.  It has also been suggested that inter-band repulsive electronic interactions may be relevant to MgB$_2$
\cite{ima01,yam01}. Below we show that the temperature dependence of the Josephson current between a multi-gap and a single-gap superconductor can
reveal the existence of $\pi$ phase difference between the order parameter on the two bands participating in the superconducting state. Such a
Josephson junction can also be used to detect a collective mode originally proposed by Leggett \cite{leg66}. This mode may exist regardless of the
relative sign of the order parameters in the two bands and involves an oscillation of their relative phase. It has been proposed theoretically for
bilayer cuprates \cite{wu95} and Sr$_2$RuO$_4$ \cite{agt97} and observed experimentally in SmLa$_{0.8}$Sr$_{0.2}$CuO$_{4-\delta}$ \cite{marel}.

In the following, we focus on a two-band superconductor with bands
labelled by $\sigma$ and $\pi$. In spin-ladder cuprates these would
correspond to the anti-bonding and bonding bands of the two legs of
the ladder. In MgB$_2$, the $\sigma$ band would correspond to the
quasi-2D hole bands due to the $\sigma$-bonding $p_{x,y}$ boron
orbitals and the $\pi$ band would correspond to the 3D electron and 3D
hole band due to the $\pi$ bonding $p_z$ boron orbitals \cite{kro01}.
To describe the superconducting state, we use a two-band BCS
model in the clean limit. This model is parameterized by the
interaction matrix that describes both the intra-band ($V_{\pi,\pi}$
and $V_{\sigma,\sigma}$) and the inter-band matrix ($V_{\sigma,\pi}$)
pair scattering elements.  In the calculations
presented below we employ the weak-coupling
self-consistency gap equation
$\Delta_{\alpha}=-\pi
T \sum_{\beta}V_{\alpha,\beta} N_{\beta}
\sum_{\omega_n}\Delta_{\beta}\,/\sqrt{\omega_n^2+|\Delta_{\beta}|^2}
$, where $N_{\beta}$ is the density of states at the Fermi surface for
band $\beta$ and $\omega_n=\pi T(2n+1)$. However we expect
our results to apply even in the strong coupling
regime.

\noindent {\it Temperature dependence of the Josephson current} A sign
difference between $\Delta_{\sigma}$ and $\Delta_{\pi}$ can only be
detected through a phase sensitive experiment. Here we examine the
Josephson current through a junction between a multi-band
superconductor and a single-band superconductor. The Josephson current
through such a junction can be found once the boundary conditions for
the quasi-classical equations have been specified.  This has been done
by Zaitsev \cite{zai84} (see also Ref.~\cite{amb63}) and generalized
by Mazin {\it et al.}  \cite{maz95} to multi-band superconductors. The
resulting current through the junction with multi-band superconductors
on both the right side ${\mathcal R}$ and the left side ${\mathcal L}$
is
\begin{eqnarray} I_S=\frac{\pi T}{e}
\sum_{i,j}\frac{1}{R_{N,ij}}\sin(\phi_{{\mathcal
L}_i}-\phi_{{\mathcal R}_j}) \times \nonumber \\
\sum_{\omega_n>0}\frac{|\Delta_{{\mathcal L}_i}(T)||\Delta_{{\mathcal R}_j}(T)|}{\sqrt{|\Delta_{{\mathcal
L}_i}(T)|^2+\omega_n^2}\sqrt{|\Delta_{{\mathcal R}_j}(T)|^2 +\omega_n^2}}
\end{eqnarray}
where $\phi_{{\mathcal R (L) }_i}$ is the phase of $\Delta_{{\mathcal
R (L)}_i}$, $R_{N,ij}^{-1}=min\{R^{-1}_{{\mathcal
L}_i},R^{-1}_{{\mathcal R}_j}\}$ with $({\mathcal A} R_{{\mathcal
L}_i({\mathcal R}_j)})^{-1}=\frac{2e^2}{\hbar}\int_{v_n>0}
\frac{D_{ij} v_{n,{\mathcal L}_i({\mathcal R}_j})d^2S_{{\mathcal L}_i
({\mathcal R}_j)}}{(2\pi)^3v_{F,{\mathcal L}_i({\mathcal R}_j)}}$,
${\mathcal A}$ is the junction area, $d^2S_{{\mathcal L}_i}$ denotes
the area element of Fermi surface ${\mathcal L}_i$, and $D_{i,j}$ is
the probability for a quasi-particle to tunnel from band $i$ in
${\mathcal L}$ to band $j$ in ${\mathcal R}$. The total junction
resistance is given by $R_N^{-1}=\sum_{ij}R_{N,ij}^{-1}$. The
functions $\Delta_{{\mathcal L}_i ({\mathcal R}_j)}(T)$ take on the
bulk values, as is justified for $s$-wave superconductors near
non-magnetic insulating surfaces.

\begin{figure}
\epsfxsize=2.5 in \center{\epsfbox{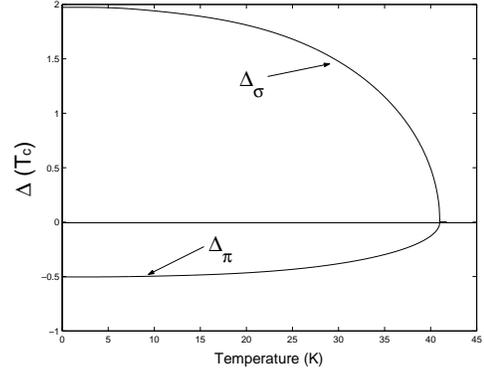}} \caption{Temperature dependence of the $\Delta_{\sigma}$ and $\Delta_{\pi}$.} \label{fig1}
\end{figure}

\begin{figure}
\epsfxsize=2.5 in \center{\epsfbox{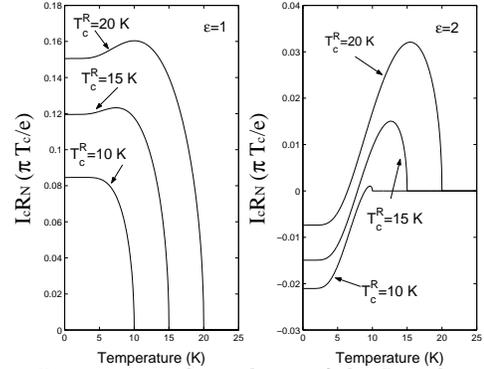}} \caption{Temperature dependence of the Josephson current between a two gap superconductor (with
$T_c=41$ K) and a single-band superconductor with various $T_c^R$ as given in the plot. The parameter $\epsilon=R_{N,\sigma}/R_{N,\pi}$. Note that
the $I_cR_N$ products for $\epsilon=2$ pass through zero to make the graph clearer, the actual values will remain positive.} \label{fig2}
\end{figure}

Here we consider the simplest case of a Josephson junction between a conventional single-band superconductor $\Delta_{\mathcal{R}}$ and a two-band
superconductor ($\Delta_{{\mathcal L},i}$, $i=\{\pi,\sigma\}$). We consider a geometry for which both bands contribute to the transport and  assume
that the conductance through the junction is limited by the two-band superconductor. In our calculations we take $V_{\pi,\pi}=0$ and
$|V_{\sigma,\pi}|=0.35 |V_{\sigma,\sigma}|$ with a density of states ratio $N_{\pi}/N_{\sigma}=1.35$ ($V_{\sigma,\pi}$ is taken negative so that
the two gaps are of opposite sign). Physically, this corresponds to a purely induced gap on the $\pi$ band (this interaction simulates that of
MgB$_2$ \cite{bou01}). We consider two possible values for $\epsilon=R_{N\sigma}/R_{N\pi}=2$ and $\epsilon=1$. Assuming a BCS theory for the
single-band superconductor we find the results shown on Figs. \ref{fig1} and ~\ref{fig2}. The most important feature for $\epsilon=1$ is that the
maximum in the Josephson current occurs at {\it finite} temperature, not at zero temperature. This maximum occurs due to thermal effects. At high
temperatures, the thermally excited quasiparticles easily deplete the Josephson current arising from the overlap of the order parameter with the
smaller amplitude in the multi-band superconductor, and the order parameter of the conventional, single band superconductor. However, as the
temperature is lowered, the contribution from the band with the smaller gap becomes more important, which leads to a downturn in the total
Josephson current. The behavior for $\epsilon=2$ is even more striking. In this case the Josephson current becomes zero at some temperature. This
remarkable behavior occurs because the $\pi$ band is assumed to have the smaller gap but the larger conductivity through the junction. The values
of $\epsilon$ that allow for a vanishing $I_c$ can be found analytically when $T_c^R\ge T_c^L$. This can be done by comparing the sign of $I_S$
given by Eq. 5 at $T=0$ and at $T=T_c^L$. If the sign changes then there must be a zero in $I_c$. For $T\approx T_c^L$,
\begin{eqnarray}
I_S=\frac{\pi F\big (|\Delta_{{\mathcal R}}(T)| \big )}{4Te}\sin(\phi_{{\mathcal L \sigma}}-\phi_{\mathcal R})\big[ \frac{|\Delta_{{\mathcal
L}_\sigma}(T)|}{R_{N,\sigma}}-\frac{|\Delta_{{\mathcal L}_\pi}(T)|}{R_{N,\pi}}\big ] \nonumber
\end{eqnarray}
where $F\big (|\Delta_{{\mathcal R}}(T)| \big )$ is a function which has no simple form (except for $T_c^R\approx T_c^L$ where $F\big
(|\Delta_{{\mathcal R}}(T)| \big )= |\Delta_{{\mathcal R}}(T)|$).  For $T=0$, \cite{amb63} and $|\Delta_{{\mathcal
L}_\sigma}(0)|>|\Delta_{{\mathcal R}}(0)|$,
\begin{eqnarray} I_S=&\sin(\phi_{{\mathcal L}_\sigma}-\phi_{{\mathcal R}}) \label{zeroT}\\&\big [\frac{1}{R_{N,\sigma}e} |\Delta_{{\mathcal
R}}(0)|K\big (\sqrt{1-(|\Delta_{{\mathcal R}}(0)|/|\Delta_{{\mathcal L}_\sigma}(0)|)^2}\big ) \nonumber \\ &-\frac{1}{R_{N,\pi}e}
|\Delta_{{\mathcal L}_\pi}(0)|K\big (\sqrt{1-(|\Delta_{{\mathcal L}_\pi}(0)|/|\Delta_{{\mathcal R}}(0)|)^2}\big )\big ] \nonumber\end{eqnarray}
where $K(x)$ is a complete elliptic integral of the first kind. For $|\Delta_{{\mathcal L}_\sigma}(0)|<|\Delta_{{\mathcal R}}(0)|$,
$|\Delta_{{\mathcal L}_\sigma}(0)|$ and $|\Delta_{{\mathcal R}}(0)|$ should be interchanged in the term proportional to $|\Delta_{{\mathcal
R}}(0)|$ in Eq.~\ref{zeroT}. For example, if $|\Delta_{\pi}|=2|\Delta_{\mathcal R}|=3|\Delta_{\sigma}|$, then the zero in $I_c$ exists for
$3.0>\epsilon>1.7$. Note that in the temperature region where the Josephson current vanishes, higher order terms in the Josephson coupling should
be included in the theory. This will not be done here.

Fig.~\ref{fig2} was determined for only one choice of the parameters
$V_{\alpha,\beta}$. We have explored a much wider parameter range
and have found that there are three criteria for the observation of
this finite temperature maximum or the vanishing of the Josephson
current: (1)- the gaps are of opposite sign; (2)- the smaller gap is
smaller than the that of the single-band superconductor (note the
larger gap can be larger or smaller than the gap of the single band
superconductor); and (3)- both bands must contribute to the
conductance through the junction (the effect always occurs for
$\epsilon=1$ when the first two conditions are satisfied). Since
superconductivity in spin ladder cuprates \cite{dag96} represents a
likely testing ground for the predicted behavior, it is worthwhile
discussing the gap structure more carefully.  In this case the
relative phase difference in $\pi$ arises from repulsive interactions
between the bonding ($\sigma$) and anti-bonding ($\pi$) bands of the
two legs of the ladder. In general, the density of states is not the
same for the $\pi$ and $\sigma$ bands \cite{nor99}. If a large on-site
Coulomb repulsion exists, then the gap structure is easily determined
by the constraint that the on-site pairing amplitude is zero. In
particular, $\sum_{k,\alpha}\Delta_{\alpha}=0$, which implies
$N_{\sigma}\Delta_{\sigma}=-N_{\pi}\Delta_{\pi}$ for all temperatures
({\it e.g.} the Fermi surface with the bigger density of states has
the smaller gap). We have confirmed that the finite temperature
maximum in $I_c$ (as seen in Fig.~\ref{fig2}) occurs for this gap
structure when $\epsilon=1$.
It is useful to point out that there have been at least three
experiments where the Josephson current exhibits a peak at finite
temperatures (similar to Fig.~\ref{fig2} for $\epsilon=1$ ): in a
UBe$_{13}$/Ta junction \cite{han86}, a YBCO/Pb junction \cite{igu94},
and a Pb/Sr$_2$RuO$_4$/Pb junction \cite{jin99}. In all three cases,
the suspected unconventional nature of UBe$_{13}$, YBCO, and
Sr$_2$RuO$_4$ have been argued to be responsible for this behavior
\cite{han86,igu94,hon98}. However, all these materials have
multiple-bands and perhaps the explanation given above is relevant.
One can also give simple generalizations of the
discussion above to the cases of more complicated junction geometries
in which different $k$-points on the Fermi surface
play the role of separate bands.

\noindent{\it Collective mode-assisted tunnelling} Here we show that Leggett's collective mode resonantly couples to the DC Josephson current of a
junction between the two-band superconductor and a single-band superconductor. The gaps need not be of opposite sign as in the previous section.
Consider a superconductor that has two superconducting order parameters $\Delta_1$ and $\Delta_2$, coming from two bands, with a Josephson coupling
between them. Let $\phi_1$ and $\phi_2$ be their respective phases. First, let us derive Leggett's mode from the microscopic equations of motion.
We define $\phi = \frac{\phi_1 - \phi_2}{2}$, the chemical potential difference between the two bands $\Delta \mu = \mu_1 - \mu_2$, and charge
imbalance between the two bands $L = \Delta Q_1 - \Delta Q_2$. They are related through an appropriate compressibility $\kappa$ and the relation $L
= \kappa \Delta \mu$. When the two bands are out of equilibrium, there is internal Josephson current $J_c \sin (2 \Delta \phi)$ and some internal
dissipative current $\sigma \Delta \mu$. From charge conservation and using the relation $[ L, \phi ] = - 2 i e$ we have
\begin{eqnarray}
\frac{\partial L}{\partial t} = -J_c \sin (2 \phi) - \sigma \Delta \mu
\\
\frac{\kappa \hbar}{e} \frac{\partial^2 \phi}{\partial t^2} = -J_c
\sin (2 \phi) - \frac{\sigma \kappa \hbar}{e} \frac{\partial
\phi}{\partial t}.
\end{eqnarray}
So, there is a collective mode at energy $\omega_0^2=2e J_c/\kappa
\hbar$ with dissipation set by $\sigma$.  As discussed in \cite{leg66} such
simplified discussion is appropriate only when $V_{\sigma,\pi}^2 <
V_{\sigma,\sigma} V_{\pi,\pi}$. The energy of the collective mode
can also be expressed using parameters of the original
microscopic Hamiltonian
$\omega_0= (\,8\, |V_{\sigma,\pi}| |\Delta_{\pi}||\Delta_{\sigma}|/[(N_{\sigma}+N_{\pi})
V_{\sigma,\sigma}V_{\pi,\pi}])^{1/2}$.
If the condition $V_{\sigma,\pi}^2 < V_{\sigma,\sigma} V_{\pi,\pi}$ is
not satisfied, there is no sharp mode but a broad continuum of
excitations corresponding to a transfer of electrons between the two
bands. For simplicity we focus the analysis in
this paper on the case when a sharp collective mode exists,
and provide a qualitative discussion of the opposite
situation, when only a broad continuum of excitations is present.

We now discuss what happens if there is Josephson coupling between
the two-gap superconductor and another single gap superconductor.
Let $\phi_3$ be the superconducting phase of this other
superconductor. The charge balance equation becomes
\begin{eqnarray}
\frac{\partial L}{\partial t} = -J_c \sin (2 \phi) - \sigma
\Delta \mu \nonumber \\ -J_1 \sin ( \phi_1 - \phi_3)+J_2 \sin
(\phi_2 - \phi_3)
\end{eqnarray}
If we introduce $\theta = \frac{\phi_1 + \phi_2}{2} - \phi_3$, we have
\begin{eqnarray}
\frac{\kappa \hbar}{e} \frac{\partial^2 \phi}{\partial t^2}
+\frac{\sigma \kappa \hbar}{e} \frac{\partial \phi}{\partial t} =
- J_c \sin (2 \phi) \nonumber \\ -J_1 \sin ( \theta + \phi) + J_2
\sin ( \theta - \phi).
\end{eqnarray}
When there is a constant voltage between the two superconductors
we have $\theta = 2eVt/\hbar=\Omega_v t$. Assuming that $\phi$ is
small, we can solve the last equation for $\phi$
\begin{eqnarray}
\phi(t) = \frac{e}{\kappa\hbar}Im
\large\{\frac{(J_1-J_2)e^{i \Omega_v t}}{[-\Omega_v^2+\omega_0^2+i\sigma
\Omega_v]} \large\}
\end{eqnarray}
If we average the total current $I_{tot}=J_1
\sin ( \phi_1 - \phi_3)+J_2 \sin ( \phi_2 - \phi_3) $ over time we find
\begin{eqnarray}
I_{tot}= \frac{e}{2\kappa\hbar}\frac{\frac{2\sigma
eV}{\hbar}(J_1-J_2)^2
}{[(\omega_0^2-\frac{4e^2V^2}{\hbar^2})^2+[\frac{2\sigma
eV}{\hbar}]^2]}
\end{eqnarray}
We therefore have a resonance enhancement of the DC current when
voltage matches the energy of the Leggett mode. If the experiments are
done at finite current this will show up as Fiske steps. There is a
simple physical interpretation of this result based on the picture of
the Leggett's mode as a bound state between quasiparticles in the two
bands \cite{leg66}. Consider a process where an electron from one
band of a
multi-gap superconductor traverses the junction, gets Andreev
reflected from the single gap superconductor and comes back as a hole
into the other band. During both crossings the quasiparticle acquires an
energy $eV$, so we end up with a pair of quasiparticles that has
energy $2 eV$. When this energy matches the energy of the bound state
$\hbar \omega_0$ this process gets resonantly enhanced and we find a
peak in the DC current.  This suggests a generalization of the above
discussion to multiple traversals of the junction by particles and
holes resulting in the creation of a Leggett exciton. In this more
general case we can expect peaks in $I(V)$ at voltages
$V_n=\hbar\omega_0/ (2 e n)$, where $n$ is any integer.  A related
scenario has been discussed recently by Auerbach and Altman
\cite{assa} to explain a subgap structure in the high T$_c$
junctions. Their proposal includes the creation of a pair of magnetic
$S=1$ excitons rather than a single $S=0$ exciton as in our case.
When the sharp resonance is absent we expect
that there will be no peak in the DC current but a jump starting from
some threshold voltage. This may be interpreted as a collection of
small peaks coming from the continuum of two quasiparticle excitations
and is similar to the usual IV characteristics for Josephson
junctions.  It is useful to point out that sharp peaks in $I(V)$ for
Josephson junctions (or Fiske steps, when the experiments are done at
a fixed current) may also arise from the Swihart waves
\cite{abrikosov}.  One should be able to separate the two mechanisms
since the energy of the resonances coming from Swihart waves depend on
the geometry of the junction, whereas the energy of the Leggett's mode
depends on the material properties only.

The interaction used in the previous section is such that the Leggett's collective mode is absent.  However, that form of interaction corresponds
to a very specific circumstance, when superconductivity in one of the bands is induced entirely by the other band. Clearly, this need not be the
case, as the Cooper instability could be present in both bands independently. Therefore, the actual interaction for the candidate multi-gap
superconductor materials are, in general, different from the simple, and rather specific model interaction we used above.  We note that an
interlayer exciton has been recently observed in the c-axis optical conductivity experiments on SmLa$_{0.8}$Sr$_{0.2}$CuO$_{4-\delta}$ by D. Dulic
{\it et al.} \cite{marel}. This mode may be understood as a particular realization of the Leggett's exciton where the two bands correspond to the
individual layers in a bilayer. Leggett's mode was  also argued to exist in MgB$_2$ and in Sr$_2$RuO$_4$ \cite{agt97}. All of these materials are
promising candidates for observing novel Josephson phenomena discussed in this paper.


\noindent{\it Conclusions} Inspired by a series of recent experimental
developments strongly supporting multi-band superconductivity in a
variety of compounds, we have examined the Josephson effect between
such multi-band superconductors and single-band superconductors. We
have shown that this can be a rather powerful probe of new physics
associated with multiple phases in the multi-band superconductors. In
particular, such junctions can be used to detect the $\pi$ phase
difference between the multiple gaps of the multi-band superconductor,
and consequently provide experimental support for the presence of
repulsive interband interactions. This would add considerable support
to the notion that an adequate description of the novel multi-band
superconductors must go beyond models considering electron-phonon
interaction alone. We have also pointed out, that such hybrid
Josephson junctions are capable of detecting collective modes arising
from the fluctuations in the relative phase between these gaps. Such
an experiment, if successful, would be the first indication that
Leggett's mode, proposed more than three decades ago, is in fact
exhibited by these novel multi-gap superconductors.

We would like to thank Prof.~A. A. Abrikosov, Dr.~G. W. Crabtree, Dr.~K. E. Gray, Dr.~D. Hinks, Dr.~W. K. Kwok, Dr.~G. Karapetrov, Prof.~J. F.
Zasadzinksi and Mr.~H. Schmidt for useful discussions. We are especially grateful to Dr.~ G. W. Crabtree and Dr.~W. K. Kwok for their hospitality
during our stay at the Materials Sciences Division of Argonne National Laboratory, where this research program was initiated. This research was
supported in part by U.S. DOE, Office of Science, under contract No. W-31-109-ENG-38. D.F.A. was also supported by the Research Corporation.

\end{document}